# A Note on the Complexity of Computing the Smallest Four-Coloring of Planar Graphs[*]

André Große  
Institut für Informatik  
Friedrich-Schiller-Universität Jena  
07740 Jena, Germany

Jörg Rothe [†]  
Institut für Informatik  
Heinrich-Heine-Universität Düsseldorf  
40225 Düsseldorf, Germany

Gerd Wechsung  
Institut für Informatik  
Friedrich-Schiller-Universität Jena  
07740 Jena, Germany

February 7, 2006

**Abstract**

We show that computing the lexicographically first four-coloring for planar graphs is $\Delta_2^p$-hard. This result optimally improves upon a result of Khuller and Vazirani who prove this problem NP-hard, and conclude that it is not self-reducible in the sense of Schnorr, assuming P $\neq$ NP. We discuss this application to non-self-reducibility and provide a general related result.

## 1 Introduction

Khuller and Vazirani [KV91] proved an NP-hardness lower bound for computing the lexicographically first solutions of the planar graph four colorability problem, which we denote by Pl-4-Color. It follows from their result that, assuming P $\neq$ NP, the polynomial-time decidable problem Pl-4-Color is not self-reducible in the sense of Schnorr [Sch76, Sch79]. Noting that their result appears to be the first such non-self-reducibility result for problems in P, they proposed as an interesting task to find other problems in P that are not self-reducible under some plausible assumption.

[*]This work was supported in part by the German Science Foundation (DFG) under grants RO 1202/9-1, RO 1202/9-3, and a Heisenberg Fellowship for the second author. A preliminary version of this paper appeared as part of [GRW01] in the proceedings of the *Seventh Italian Conference on Theoretical Computer Science*.

[†]Corresponding author. Email: rothe@cs.uni-duesseldorf.de.

In this note, we raise Khuller and Vazirani's NP-hardness lower bound for computing the lexicographically smallest four-coloring of a planar graph to $\Delta_2^p$-hardness. Our result is optimal, since this problem belongs to (the function analog of) the class $\Delta_2^p$.

The class $\Delta_2^p = \text{P}^{\text{NP}}$, which belongs to the second level of the polynomial hierarchy [MS72, Sto77], contains exactly the problems solvable in deterministic polynomial time with an NP oracle. Papadimitriou [Pap84] proved that Unique-Optimal-Traveling-Salesperson is $\Delta_2^p$-complete, and Krentel [Kre88] and Wagner [Wag87] established many more $\Delta_2^p$-completeness results, including the result that the problem Odd-Max-SAT is $\Delta_2^p$-complete. The complexity of colorability problems has been studied in a number of papers, see, e.g., [AH77a, AH77b, Sto73, GJS76, Wag87, KV91, Rot03].

As mentioned above, if for some problem in P computing the lexicographically smallest solution is hard, then the problem itself cannot be self-reducible in the sense of Schnorr [Sch76, Sch79], unless $\text{P} = \text{NP}$. We discuss this application to non-self-reducibility and provide a general related result. In particular, it follows from this result that even a set as simple as $\Sigma^*$ has representations in which it is not self-reducible in Schnorr's sense, unless $\text{P} = \text{NP}$.

## 2 Computing the Smallest Four-Coloring of a Planar Graph

Solving the famous Four Color Conjecture in the affirmative, Appel and Haken [AH77a, AH77b] showed that every planar graph can be colored with no more than four colors. In contrast, for each $k \geq 4$, computing the lexicographically first $k$-coloring of a planar graph is hard: Khuller and Vazirani [KV91] established an NP-hardness lower bound for this problem. We raise their lower bound to $\Delta_2^p$-hardness. Since the lexicographically smallest $k$-coloring of a planar graph can be computed in (the function analog of) $\Delta_2^p$, this improved lower bound is optimal.

**Definition 2.1** *Let $k > 1$, and let $0, 1, \ldots, k-1$ represent $k$ colors.*

- *A $k$-coloring of an undirected graph $G = (V, E)$ is a mapping $\psi_G : V \to \{0, 1, \ldots, k-1\}$.*

- *A $k$-coloring $\psi_G$ is said to be* legal *if and only if for each edge $\{u, v\} \in E$, $\psi_G(u) \neq \psi_G(v)$.*

- *A graph $G$ is said to be $k$-colorable if and only if there exists a legal $k$-coloring of $G$.*

- *Let Pl-$k$-Color denote the planar graph $k$-colorability problem.*

Stockmeyer [Sto73] proved that Pl-3-Color is NP-complete, see also Garey et al. [GJS76]. By Appel and Haken's above-mentioned result, every planar graph is four-colorable. Thus, Pl-$k$-Color is in P for each $k \geq 4$.

**Definition 2.2 (Khuller and Vazirani [KV91])** *Let $k > 1$, and let the vertex set of a given undirected graph $G = (V, E)$ with $n$ vertices be ordered as $V = \{v_1, v_2, \ldots, v_n\}$. Then, every $k$-coloring $\psi_G$ of $G$ can be represented by a string $\psi_G$ in $\{0, 1, \ldots, k-1\}^n$:*

$$\psi_G = \psi_G(v_1)\psi_G(v_2)\cdots\psi_G(v_n).$$

*Define the* lexicographically smallest (legal) $k$-coloring *by*

$$\text{LF}_{\text{Pl-}k\text{-Color}}(G) = \min\{\psi_G \mid \psi_G \text{ is a legal } k\text{-coloring of } G\},$$

*if $G \in \text{Pl-}k\text{-Color}$, where the minimum is taken with respect to the lexicographic ordering of strings, and define* $\text{LF}_{\text{Pl-}k\text{-Color}}(G) = 10^n$ *if $G \notin \text{Pl-}k\text{-Color}$.*

We now prove our main result.

**Theorem 2.3** *Computing the lexicographically smallest $k$-coloring for planar graphs is $\Delta_2^p$-hard for any $k \geq 4$.*

**Proof.** For simplicity, we show this claim only for $k = 4$. Let $\rho_4$ be the reduction of Khuller and Vazirani [KV91, Theorem 3.1]. Recall that $\rho_4$ maps a given planar graph $G = (V, E)$, whose vertices are ordered as $V = \{v_1, v_2, \ldots, v_m\}$, to the planar graph $H = (U, F)$ defined as follows:

- The vertex set of $H$ is ordered as $U = \{u_1, u_2, \ldots, u_{2m}\}$, where $u_i$ is a new vertex and $u_{m+i} = v_i$ is an old vertex for each $i$, $1 \leq i \leq m$.

- The edge set of $H$ is defined by $F = E \cup \{\{u_i, u_{m+i}\} \mid 1 \leq i \leq m\}$.

It follows immediately from this construction that

(1) $\qquad G \in \text{Pl-3-Color} \iff \text{LF}_{\text{Pl-4-Color}}(\rho_4(G)) \in \{0^m w \mid w \in \{1,2,3\}^m\},$

that is, "$G \in \text{Pl-3-Color}$?" can be decided by looking at the first $m$ bits of $\text{LF}_{\text{Pl-4-Color}}(H)$.

We give a reduction from the problem Odd-Min-SAT, which is defined to be the set of all boolean formulas $F = F(x_1, x_2, \ldots, x_n)$ in conjunctive normal form for which, assuming $F$ *is* satisfiable, the lexicographically smallest satisfying assignment $\alpha : \{x_1, x_2, \ldots, x_n\} \to \{1, 2\}$ is "odd," i.e., for which $\alpha(x_n) = 1$. Here, "1" represents "true," and "2" represents "false."

It is well known that Odd-Min-SAT is $\Delta_2^p$-complete; Krentel [Kre88] and also Wagner [Wag87] proved the corresponding claim for the dual problem Odd-Max-SAT.

Let $F = F(x_1, x_2, \ldots, x_n)$ be any given boolean formula. Without loss of generality, we may assume that $F$ is in conjunctive normal form with exactly three literals per clause. Assume that $F$ has $z$ clauses. Let $\sigma$ be the Stockmeyer reduction from 3-SAT to Pl-3-Color, see Stockmeyer [Sto73] and also Garey et al. [GJS76]. This reduction $\sigma$, on input $F$, yields a graph $G = (V, E)$ with $m > n$ vertices, where $m = m(F)$ depends on the number $n$ of variables, the number $z$ of clauses, and the structure of $F$. Note that $F$'s structure induces a certain number of "crossovers" of edges to guarantee the planarity of $G$; see [GJS76, Sto73] for details.

Order the vertex set of $G$ as $V = \{v_1, v_2, \ldots, v_m\}$ such that

(a) for each $i$, $1 \leq i \leq n$, $v_i$ represents the variable $x_i$, and

(b) for each $i$, $n < i \leq m$, $v_i$ represents some other vertex of $G$.

Note that $G$ is a planar graph satisfying that

(i) $F$ is satisfiable if and only if $G$ is 3-colorable, using the colors 1, 2, and 3, and

(ii) every satisfying assignment $\alpha$ of $F$ corresponds to a 3-coloring $\psi_\alpha$ of $G$ such that for each $i$, $1 \leq i \leq n$, $\psi_\alpha(v_i) = \alpha(v_i) \in \{1, 2\}$. The color 3 is used for the other vertices of $G$.

Now apply the reduction $\rho_4$ of Khuller and Vazirani to $G$ and obtain a planar graph $H = \rho_4(G) = \rho_4(\sigma(F))$ that satisfies Equation (1) as described above. It follows immediately from this construction and from Equation (1) that

$$F \in \texttt{Odd-Min-SAT} \iff$$
$$\text{LF}_{\texttt{Pl-4-Color}}(\rho_4(\sigma(F))) \in \{0^m w 1 y \,|\, w \in \{1,2\}^{n-1} \text{ and } y \in \{1,2,3\}^{m-n}\},$$

that is, "$F \in \texttt{Odd-Min-SAT}$?" can be decided by looking at the first $m$ bits and at the $(m+n)$th bit of $\text{LF}_{\texttt{Pl-4-Color}}(H)$.

For $k > 4$, the claim of the theorem follows from an analogous argument that employs in place of $\rho_4$ the appropriate reduction $\rho_k$ from [KV91, Thm. 3.2]. ∎

## 3 Non-Self-Reducible Sets in P

From their NP-hardness lower bound for computing the lexicographically first four-coloring of planar graphs, Khuller and Vazirani [KV91] conclude that, unless $\text{P} = \text{NP}$, the polynomial-time decidable problem $\texttt{Pl-}k\texttt{-Color}$ is not self-reducible for $k \geq 4$. The type of (functional) self-reducibility used by Khuller and Vazirani is due to Schnorr [Sch76, Sch79], see also [BD76]. For more background on self-reducibility, see, e.g., [Sel88, JY90, Rot05].

**Definition 3.1 (Schnorr [Sch76, Sch79])**

- *Let $\Sigma$ and $\Gamma$ be alphabets with at least two symbols each. Instances of problems are encoded over $\Sigma$, and solutions of problems are encoded over $\Gamma$. For any set $B \subseteq \Sigma^* \times \Gamma^*$ and any polynomial $p$, the $p$-projection of $B$ is defined to be the set*

$$\text{proj}_p(B) = \{x \in \Sigma^* \,|\, (\exists y \in \Gamma^*)\,[|y| \leq p(|x|) \text{ and } (x,y) \in B]\}.$$

- *A partial order $\leq$ on $\Sigma^*$ is* polynomially well-founded and length-bounded *if and only if there exists a polynomial $q$ such that*

  (a) *every $\leq$-decreasing chain with maximum element $x$ has at most $q(|x|)$ elements, and*

  (b) *for all strings $x, y \in \Sigma^*$, $x < y$ implies $|x| \leq q(|y|)$.*

- *Let $A = \text{proj}_p(B)$ for some set $B \subseteq \Sigma^* \times \Gamma^*$ and some polynomial $p$. The projection $A$ is said to be* self-reducible with respect to $(B, p)$ *if and only if there exist a polynomial-time computable function $g$ mapping from $\Sigma^* \times \Gamma$ to $\Sigma^*$ and a polynomially well-founded and length-bounded partial order $\leq$ such that for all strings $x \in \Sigma^*$, for all strings $y \in \Gamma^*$, and for all symbols $\gamma \in \Gamma$,*

(i) $g(x, \gamma) < x$, and

(ii) $(x, \gamma y) \in B \iff (g(x, \gamma), y) \in B$.

*If the pair $(B, p)$ for which $A = \text{proj}_p(B)$ is clear from the context, we omit the phrase "with respect to $(B, p)$."*

We mention in passing that various other important types of self-reducibility have been studied, such as the self-reducibility defined by Meyer and Paterson [MP79] and the disjunctive self-reducibility studied by Selman [Sel88], Ko [Ko83], and many others. We refer the reader to the excellent survey by Joseph and Young [JY90] for an overview and for pointers to the literature. Note that, in sharp contrast with Schnorr's self-reducibility, every set in P is self-reducible in the sense of Meyer and Paterson [MP79], Ko [Ko83], and Selman [Sel88].

**Definition 3.2** *Let $\Sigma = \{0, 1\}$. Given any set $A$ in NP with $A \subseteq \Sigma^*$, there is an associated set $B_A \subseteq \Sigma^* \times \Sigma^*$ and an associated polynomial $p_A$ such that $B_A$ is in P and $A = \text{proj}_{p_A}(B_A)$.*

- *For any $x \in \Sigma^*$, define the* set of solutions *for $x$ with respect to $B_A$ and $p_A$ by*

$$\text{Sol}_{(B_A, p_A)}(x) = \{y \in \Sigma^* \mid |y| \leq p_A(|x|) \text{ and } (x, y) \in B_A\}.$$

*Note that $x \in A$ if and only if $\text{Sol}_{(B_A, p_A)}(x) \neq \emptyset$.*

- *For any $x \in \Sigma^*$, define the* lexicographically first solution *with respect to $B_A$ and $p_A$ by*

$$\text{LF}_{(B_A, p_A)}(x) = \begin{cases} \min \text{Sol}_{(B_A, p_A)}(x) & \text{if } x \in A \\ \text{bin}(2^{p(|x|)}) & \text{otherwise,} \end{cases}$$

*where the minimum is taken with respect to the lexicographic ordering of $\Sigma^*$, and $\text{bin}(n)$ denotes the binary representation of the integer $n$ without leading zeros.*

*If the pair $(B_A, p_A)$ for which $A = \text{proj}_{p_A}(B_A)$ is clear from the context, we use $\text{Sol}_A(x)$ and $\text{LF}_A(x)$ as shorthands for respectively $\text{Sol}_{(B_A, p_A)}(x)$ and $\text{LF}_{(B_A, p_A)}(x)$.*

It is well known that if $A$ is self-reducible then $\text{LF}_A$ can be computed in polynomial time by prefix search, via suitable queries to the oracle $A$. Moreover, if $A$ is in P then $\text{LF}_A$ can even be computed in polynomial time without any oracle queries. It follows that if computing $\text{LF}_A$ is NP-hard then $A$ cannot be self-reducible, assuming $\text{P} \neq \text{NP}$.

Khuller and Vazirani [KV91] propose to prove polynomial-time decidable problems other than Pl-4-Color non-self-reducible, under the assumption $\text{P} \neq \text{NP}$. As Theorem 3.5 below, we provide a general result showing that it is almost trivial to find such problems: For any NP problem $A$ for which $\text{LF}_A$ is hard to compute, one can define a P-decidable version $D$ of $A$ such that $\text{LF}_D$ is still hard to compute; hence, $D$ is not self-reducible, assuming $\text{P} \neq \text{NP}$.

To formulate this result, we now define the functional many-one reducibility that was introduced by Vollmer [Vol94] as a stricter reducibility notion than Krentel's metric reducibility [Kre88]. We also define the function class $\min \cdot \text{P}$ that was introduced by Hempel and Wechsung [HW00].

**Definition 3.3 (Vollmer [Vol94])** *Let $f$ and $h$ be functions from $\Sigma^*$ to $\Sigma^*$.*

- *We say $f$ is* polynomial-time functionally many-one reducible *to $h$ (in symbols, $f \leq_m^{\mathrm{FP}} h$) if and only if there exists a polynomial-time computable function $g$ such that for all $x \in \Sigma^*$, $f(x) = h(g(x))$.*

- *We say $h$ is $\leq_m^{\mathrm{FP}}$-hard for a function class $\mathcal{C}$ if and only if for every $f \in \mathcal{C}$, $f \leq_m^{\mathrm{FP}} h$.*

- *We say $h$ is $\leq_m^{\mathrm{FP}}$-complete for $\mathcal{C}$ if and only if $h \in \mathcal{C}$ and $h$ is $\leq_m^{\mathrm{FP}}$-hard.*

**Definition 3.4 (Hempel and Wechsung [HW00])** *Define the class $\min \cdot \mathrm{P}$ to consist of all functions $f$ for which there exist a set $A \in \mathrm{P}$ and a polynomial $p$ such that for all $x \in \Sigma^*$,*

$$f(x) = \min\{y \in \{0,1\}^* \mid |y| \leq p(|x|) \text{ and } \langle x, y \rangle \in A\},$$

*where $\langle \cdot, \cdot \rangle : \Sigma^* \times \Sigma^* \to \Sigma^*$ is a standard pairing function. If the set over which the minimum is taken is empty, define by convention $f(x) = \mathrm{bin}(2^{p(|x|)})$.*

Note that $\mathrm{LF}_A = \mathrm{LF}_{(B,p)}$ is in $\min \cdot \mathrm{P}$ for every NP set $A$ and for every representation of $A$ as a $p$-projection $A = \mathrm{proj}_p(B)$ of some suitable set $B \in \mathrm{P}$ and polynomial $p$.

**Theorem 3.5** *Let $A$ be any set in* NP*, and let $B$ and $D$ be sets in* P *and $p$ be a polynomial such that $A = \mathrm{proj}_p(B) \subseteq D$ and $\mathrm{LF}_A$ is $\leq_m^{\mathrm{FP}}$-complete for $\min \cdot \mathrm{P}$. Then, there exist a set $C \in \mathrm{P}$ and a polynomial $q$ such that $D = \mathrm{proj}_q(C)$ and computing $\mathrm{LF}_D$ is $\Delta_2^p$-hard.*

*Hence, $D$ is not self-reducible with respect to $(C, q)$, assuming $\mathrm{P} \neq \mathrm{NP}$.*

**Proof.** Let $A$, $B$, and $p$ be given as in the theorem, where $A \subseteq \Sigma^*$ and $B \subseteq \Sigma^* \times \Sigma^*$ and $\Sigma = \{0, 1\}$. Let $D$ be any set in P with $A \subseteq D$. Define

$$C = B \cup \{(x, \mathrm{bin}(2^{p(|x|)})) \mid x \in D\},$$

and let $q(n) = p(n) + 1$ for all $n$. Note that $C \in \mathrm{P}$ and $D = \mathrm{proj}_q(C)$. It also follows that $\mathrm{LF}_D(x) = \mathrm{LF}_A(x)$ if $x \in D$, and $\mathrm{LF}_D(x) = 2 \cdot \mathrm{LF}_A(x)$ if $x \notin D$.

We now show that computing $\mathrm{LF}_D$ is as hard as deciding the $\Delta_2^p$-complete problem Odd-Min-SAT, which was defined in Section 2. Since $\mathrm{LF}_A$ is $\leq_m^{\mathrm{FP}}$-complete for $\min \cdot \mathrm{P}$, we have $\mathrm{LF}_{\mathtt{SAT}}(F) = \mathrm{LF}_A(t(F))$ for some polynomial-time computable function $t$. Hence,

$$\begin{aligned} F \in \mathtt{Odd\text{-}Min\text{-}SAT} &\iff \mathrm{LF}_{\mathtt{SAT}}(F) \equiv 1 \bmod 2 \\ &\iff \mathrm{LF}_A(t(F)) \equiv 1 \bmod 2 \\ &\iff \mathrm{LF}_D(t(F)) \equiv 1 \bmod 2. \end{aligned}$$

Thus, one can decide "$F \in \mathtt{Odd\text{-}Min\text{-}SAT}$?" by looking at the last bit of $\mathrm{LF}_D(t(F))$. ∎

**Corollary 3.6** *If $\mathrm{P} \neq \mathrm{NP}$ then $\Sigma^*$ has representations in which it is not self-reducible.*

**Proof.** Replacing the set $D$ of Theorem 3.5 by $\Sigma^*$, it is clear that the hypothesis of the theorem can be satisfied by suitably choosing $A$, $B$, and $p$. It follows that $\Sigma^*$, unconditionally, has representations in which it is not self-reducible in the sense of Schnorr, unless $\mathrm{P} = \mathrm{NP}$. ∎

# 4 Conclusions and Open Questions

In Theorem 2.3, we strengthened Khuller and Vazirani's [KV91] lower bound for computing the lexicographically first four-coloring for planar graphs from NP-hardness to $\Delta_2^p$-hardness. The non-self-reducibility of the Pl-4-Color problem follows immediately from these lower bounds. Khuller and Vazirani [KV91] asked whether similar non-self-reducibility results can be proven for problems in P other than Pl-4-Color, under some plausible assumption such as P $\neq$ NP. We established as Theorem 3.5 a general result showing that it is almost trivial to find such problems.

This general result subsumes a number of results [Gro99] providing concrete—although somewhat artificial—problems in P that are not self-reducible in Schnorr's sense, unless P = NP. Why are these problems artificial? The reason is that they are P versions of standard NP-complete problems—such as the satisfiability problem, the clique problem, and the knapsack problem—that are defined by

(a) encoding directly into each solvable problem instance a trivial solution to this instance, and simultaneously

(b) ensuring that computing the smallest solution remains a hard problem by fixing a suitable ordering of the solutions to a given problem instance.

Here are some examples of such problems:

1. (a) P-SAT is the set of pairs $\langle F, x_i \rangle$ such that $F$ is a boolean formula in conjunctive normal form and $x_i$ is a variable occurring in each clause of $F$ in positive form.

   (b) Let the variables of a given formula $F$ be ordered as $F = F(x_1, x_2, \ldots, x_n)$. Just as for the satisfiability problem, a *solution to a P-SAT instance* $I = \langle F, x_i \rangle$ is any satisfying assignment $\psi_I$ of $F$. A solution $\psi_I$ of $I$ is represented by the string $\psi_I = \psi_I(x_1)\psi_I(x_2)\cdots\psi_I(x_n)$ from $\{0,1\}^n$, where "1" represents "true" and "0" represents "false."

2. (a) P-Clique is the set of pairs $\langle G, C \rangle$ such that $G = (V, E)$ is a graph and $C \subseteq V$ is a clique in $G$.

   (b) Let the vertex set of a given graph $G = (V, E)$ be ordered as $V = \{v_1, v_2, \ldots, v_n\}$. Just as for the clique problem, a *solution to a P-Clique instance* $I = \langle G, C \rangle$ is any clique $\hat{C} \subseteq V$ that is of size at least $||C||$. A solution $\hat{C}$ of $I$ is represented by the string $\psi_I = \chi_{\hat{C}}(v_1)\chi_{\hat{C}}(v_2)\cdots\chi_{\hat{C}}(v_n)$ from $\{0,1\}^n$, where $\chi_{\hat{C}}$ denotes the characteristic function of $\hat{C}$, i.e., $\chi_{\hat{C}}(v) = 1$ if $v \in \hat{C}$, and $\chi_{\hat{C}}(v) = 0$ if $v \notin \hat{C}$.

3. (a) P-Knapsack is the set of tuples $\langle U, s, v, k, b \rangle$ such that $U$ is a finite set, $s$ and $v$ are functions mapping from $U$ to the positive integers, and there exists an element $u \in U$ satisfying $s(u) \leq b$ and $v(u) \geq k$.

   (b) Let the set $U$ of a given P-Knapsack instance $I = \langle U, s, v, k, b \rangle$ be ordered as $U = \{u_1, u_2, \ldots, u_n\}$. Just as for the knapsack problem, a *solution to I* is any subset $\hat{U} \subseteq U$ that satisfies the "knapsack property," i.e., that satisfies the conditions

$$\sum_{u \in \hat{U}} s(u) \leq b \quad \text{and} \quad \sum_{u \in \hat{U}} v(u) \geq k.$$

A solution $\hat{U}$ of $I$ is represented by the string $\psi_I = \chi_{\hat{U}}(v_1)\chi_{\hat{U}}(v_2)\cdots\chi_{\hat{U}}(v_n)$ from $\{0,1\}^n$.

Note that the lexicographic ordering of strings induces a suitable ordering of the solutions to a given problem instance. For each of the above-defined problems $\Pi \in \mathrm{P}$, computing $\mathrm{LF}_\Pi$ can be shown to be NP-hard [Gro99], which implies that $\Pi$ is non-self-reducible unless $\mathrm{P} = \mathrm{NP}$.

Analogously, every standard NP-complete problem yields such an artificial, non-self-reducible problem in P. In contrast, the Pl-4-Color problem is a quite natural problem. Is it possible to prove, under a plausible assumption such as $\mathrm{P} \neq \mathrm{NP}$, the non-self-reducibility of other *natural* problems in P?

**Acknowledgments.** We thank the anonymous referees of the conference version of this paper for their helpful and insightful comments on the paper. In particular, we thank the referee who suggested an idea that led to Theorem 3.5, which subsumes some results from an earlier draft of this paper.